\documentclass[preprint,endfloats,showpacs,preprintnumbers,prl,floatfix]{revtex4}

\usepackage{graphicx}
\usepackage{epsfig}
\usepackage{bm}
\usepackage{amssymb,amsmath}
\def\4he{$^4$He}
\def\3he{$^3$He}
\def\cm3{cm$^{-3}$}
\def\cmps{cm$^3$~s$^{-1}$}

\begin{document}

\title{Large spin relaxation rates in trapped submerged-shell atoms}

\author{Colin~B.~Connolly,$^{1,3}$ Yat~Shan~Au,$^{1,3}$ S.~Charles~Doret,$^{1,3}$ Wolfgang~Ketterle,$^{2,3}$ and John~M.~Doyle$^{1,3}$}
\affiliation{
$^1$Department of Physics, Harvard University,  Cambridge, Massachusetts 02138\\
$^2$Department of Physics, MIT, Cambridge, Massachusetts 02139\\
$^3$Harvard-MIT Center for Ultracold Atoms, Cambridge, Massachusetts 02138
}
\date{\today}

\begin{abstract}
Spin relaxation due to atom--atom collisions is measured for magnetically trapped erbium and thulium atoms at a temperature near 500~mK.  The rate constants for Er--Er and Tm--Tm collisions are $3.0\times 10^{-10}$~{\cmps} and $1.1\times 10^{-10}$~{\cmps}, respectively, 2--3 orders of magnitude larger than those observed for highly magnetic $S$-state atoms.  This is strong evidence for an additional, dominant, spin relaxation mechanism, electrostatic anisotropy, in collisions between these ``submerged-shell" $L\neq0$ atoms.  These large spin relaxation rates imply that evaporative cooling of these atoms in a magnetic trap will be highly inefficient.
\end{abstract}

\pacs{34.50.-s,37.10.De,32.70.Jz}


\maketitle


Research in cold and ultracold atoms has in recent years increasingly broadened scope beyond the alkali atoms to explore and exploit the diverse range of atomic and chemical properties found across the periodic table.  In particular, the lanthanide rare-earth (RE) atoms have attracted considerable experimental and theoretical interest.  Recent experiments with RE atoms have resulted in Bose-Einstein condensation of Yb \cite{Takasu03etal}, magneto-optical trapping of Er \cite{McClelland06} and Dy \cite{YounUnpub}, Zeeman slowing of Tm \cite{Chebakov09etal}, and large ensembles ($>10^{11}$ atoms) of buffer-gas loaded and magnetically trapped RE atoms of several species below 1~K \cite{Kim97etal, Hancox04etal}.  This interest in RE systems stems from important, sometimes unique, attributes such as narrow transitions which allow for low Doppler cooling limits and improved frequency standards \cite{Berglund07, Hall89}, large magnetic moments with strong long-range dipolar interactions, and ``submerged-shell" character that in certain circumstances can shield atom--atom interactions from anisotropic valence electron shells \cite{Hancox04etal}.  Progress with these systems---or any novel atomic system---is dependent on collisional processes, in particular low rates of inelastic collisions including spin relaxation collisions in trapped samples.  Spin relaxation can cause heating as well as drive atoms out of the desired quantum state, thus preventing cooling to lower temperatures and limiting experimental sensitivity and the capacity for new discovery. 

Previous experiments and theoretical work with RE atoms, including Er and Tm, discovered suppression of electrostatic anisotropy in RE--helium collisions.  Specifically, in this interaction the anisotropic $4f$ electron distribution was found to be shielded by closed $5s$ and $6s$ electron shells \cite{Hancox04etal, KremsRE05, BuchachenkoPot06, BuchachenkoRate06etal, Chu07}.  This ``submerged-shell" nature allowed for sympathetic cooling of RE atoms by cold He and efficient buffer-gas trapping of large numbers of atoms ($>10^{11}$) at millikelvin temperatures.  It also explained reduced collisional frequency broadening of hyperfine clock transitions in Tm \cite{Aleksandrov83etal}.  The discovery of efficient shielding and consequent low inelastic rates in the RE--He system gave hope that similar suppression would be found in RE--RE collisions and could allow for efficient evaporative cooling \cite{Ketterle96}.  This could, for example, provide a path to quantum degeneracy for magnetically trapped RE atoms.

In this Letter, we present measurements of spin relaxation rates in two-body collisions of trapped submerged-shell species Er ($L=5$) and Tm ($L=3$), finding them to be very large, in striking contrast to the low rates observed in RE--He systems.  These large rates imply an additional spin relaxation mechanism other than spin exchange, dipolar relaxation, and second-order spin-orbit coupling, which are well-known from studies with alkali atoms.  Although theory has proven very effective for understanding the RE--He system, the current theory of RE--RE cold collisions is incomplete.  Despite the development by Krems et al.~of a theoretical framework for collisions of two $L\neq0$ atoms \cite{Krems04}, in the RE case these calculations are extremely difficult and to our knowledge no theoretical predictions yet exist.  Recently, an experiment was done with the transition metal titanium at a temperature of 5~K \cite{Lu09}.  Ti has submerged-shell structure, as observed in Ti--He collisions \cite{HancoxTi05etal, KremsTrans05etal}, but an order of magnitude weaker shielding of electrostatic anisotropy \cite{BuchachenkoPot06}.  Rapid decay of $^{50}$Ti electron spin polarization was observed due to $^{50}$Ti--Ti collisions, but the mechanism of this loss could not be determined because spin relaxation could not be separated from spin exchange with unpolarized Ti isotopes.  Thus, whether or not submerged-shell atoms exhibit low spin relaxation rates---indicating submerged-shell behavior in these processes---had remained an open question.


\begin{figure}
\includegraphics[width=8.0 cm]{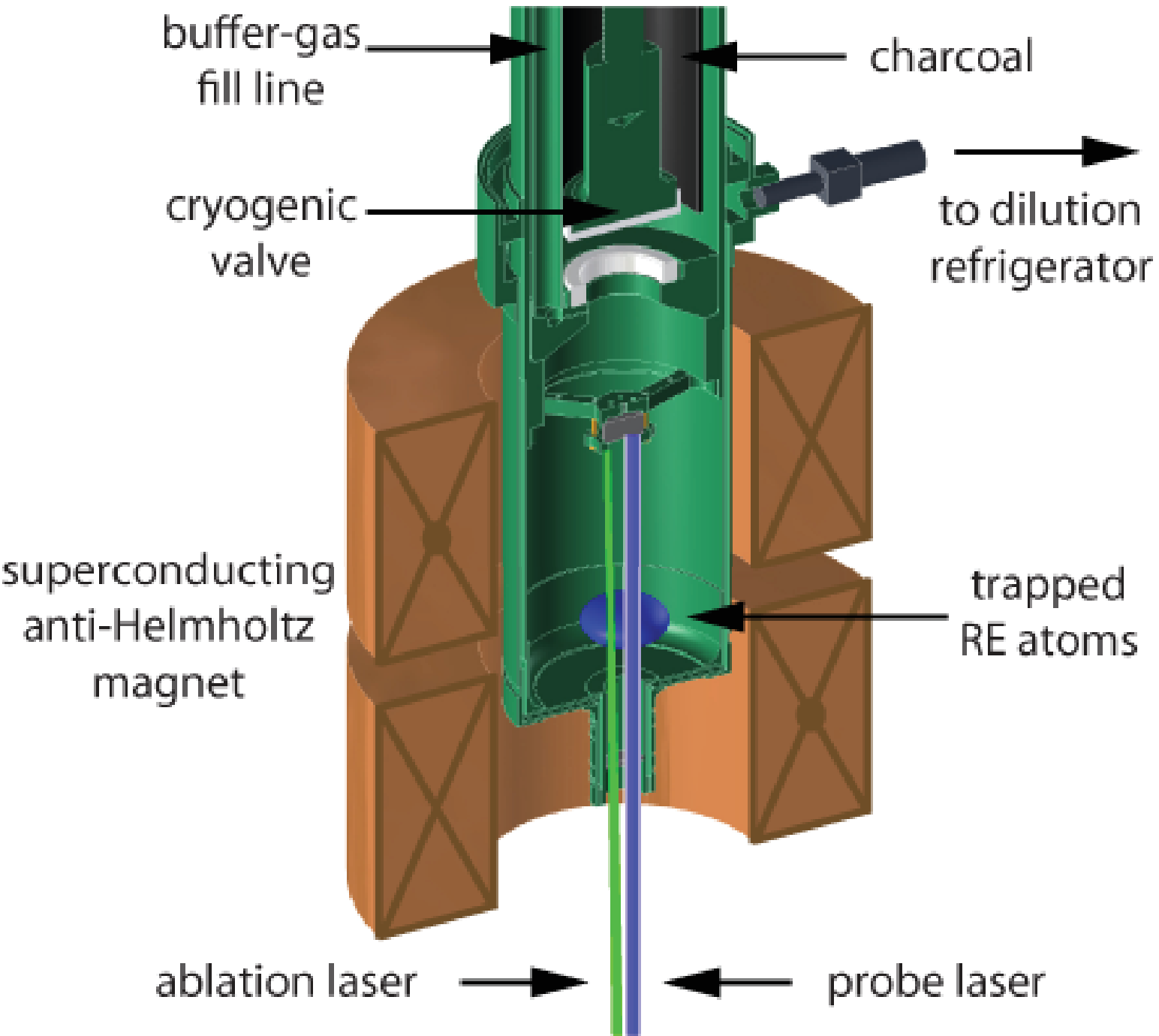}
\caption{\label{fig:apparatus}(color online) Diagram of the buffer-gas trapping apparatus.  A cryogenic valve separates the trapping region of the experimental cell from an additional pumping region and can be used to regulate the buffer-gas density.}
\end{figure}


Our experiment takes place in a double-walled plastic cell maintained at a temperature of ${\approx500}$~mK by a superfluid helium heat link to a dilution refrigerator (see Fig.~\ref{fig:apparatus}).  We produce either trapped atomic Er or Tm by laser ablation of solid metal foils into \4he buffer-gas in the presence of a magnetic quadrupole field (trap depth up to $3.7$~T) produced by large superconducting anti-Helmholtz coils surrounding the cell.  The ablated atoms cool via elastic collisions with the cold buffer-gas and within $50$~ms \cite{HancoxThesis} assume a Boltzmann distribution in the trap with peak density of up to $7\times10^{11}$~\cm3.  The trapped cloud is interrogated via laser absorption spectroscopy on the $400.9$~nm ($J=6\rightarrow7$) and $415.2$~nm ($J=6\rightarrow5$) transitions of Er and on the $409.5$~nm ($J=7/2\rightarrow5/2$) transition of Tm.


The amount of buffer-gas in the cell is regulated such that the He density is sufficient to cool the atoms after ablation, but insufficient to cause significant atom loss from RE--He collisions.  This regulation is accomplished by independent control of the cell temperature and the amount of He initially present in the cell.  The lack of observed loss from buffer-gas collisions 1~second after ablation implies a He density less than $10^{12}$~\cm3 \cite{HancoxThesis}.  Since the observed ablation yield implies a higher initial buffer-gas density, it is likely that heating from the ${\sim5}$~mJ ablation pulse temporarily desorbs additional He from the cell walls, which re-adsorbs rapidly.  We deliberately maintain a buffer-gas density of ${\approx10^{11}}$~\cm3 after trap loading in order to maintain thermal equilibrium between the trapped atoms and the cell.

%


\begin{figure}
\includegraphics[width=8.0 cm]{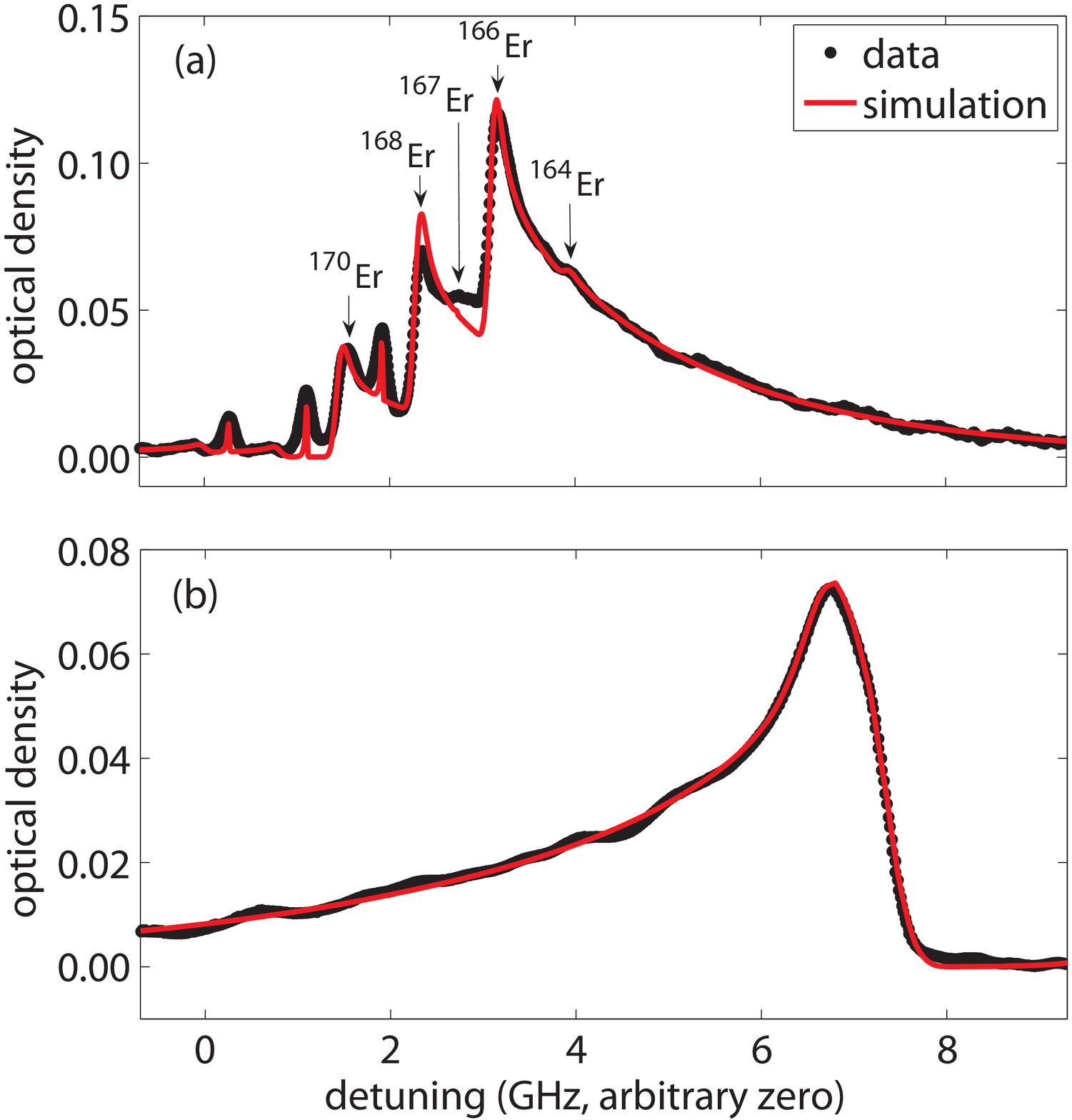}
\caption{\label{fig:spectrum}(color online) (a) Absorption spectrum of Er on the $400.9$~nm ($J=6\rightarrow7$) transition in a $0.99$~T ($4.6$~K) deep magnetic trap at 530~mK with peak density of $4.6\times10^{10}$~\cm3.  The $\Delta m_J=+1$ magnetically broadened peaks of the dominant isotopes are labeled.  The sharper peaks are $\Delta m_J=0$ transitions. Hyperfine constants are unknown for the $^{167}$Er isotope ($23\%$ abundance), and it is ignored in the spectrum simulation.  Due to the substantial Zeeman broadening, this does not significantly affect the implied atom density and temperature. (b) Absorption spectrum of Tm on the $409.5$~nm ($J=7/2\rightarrow5/2$) transition in a $3.3$~T ($8.8$~K) deep trap at 500~mK with peak density of $3.8\times10^{11}$~\cm3.  Tm has a single isotope with $I=1/2$ and known hyperfine splitting \cite{Akimov08etal}.}
\end{figure}



Example spectra of magnetically trapped Er and Tm are shown in Fig.~\ref{fig:spectrum}, showing peaks for both Zeeman-broadened $\Delta m_J=\pm 1$ and narrow $\Delta m_J=0$ transitions.  The relative magnitudes of spectral features may be used to estimate the $m_J$ state distribution, however for the case of a probe laser passing through the trap center the absorption is primarily determined by the total peak atom density rather than the contributions from individual states.  Isotope shifts for the $400.9$~nm transition of Er were not found in the literature and were determined for nuclear spin-zero isotopes by fitting to spectra measured in zero field at ${\sim4}$~K.  The shifts for isotopes $^{164}$Er, $^{168}$Er, and $^{170}$Er from the $^{166}$Er peak are $-0.80(4)$~GHz, $0.81(1)$~GHz, and $1.66(2)$~GHz, respectively.

\begin{figure}
\includegraphics[width=8.0 cm]{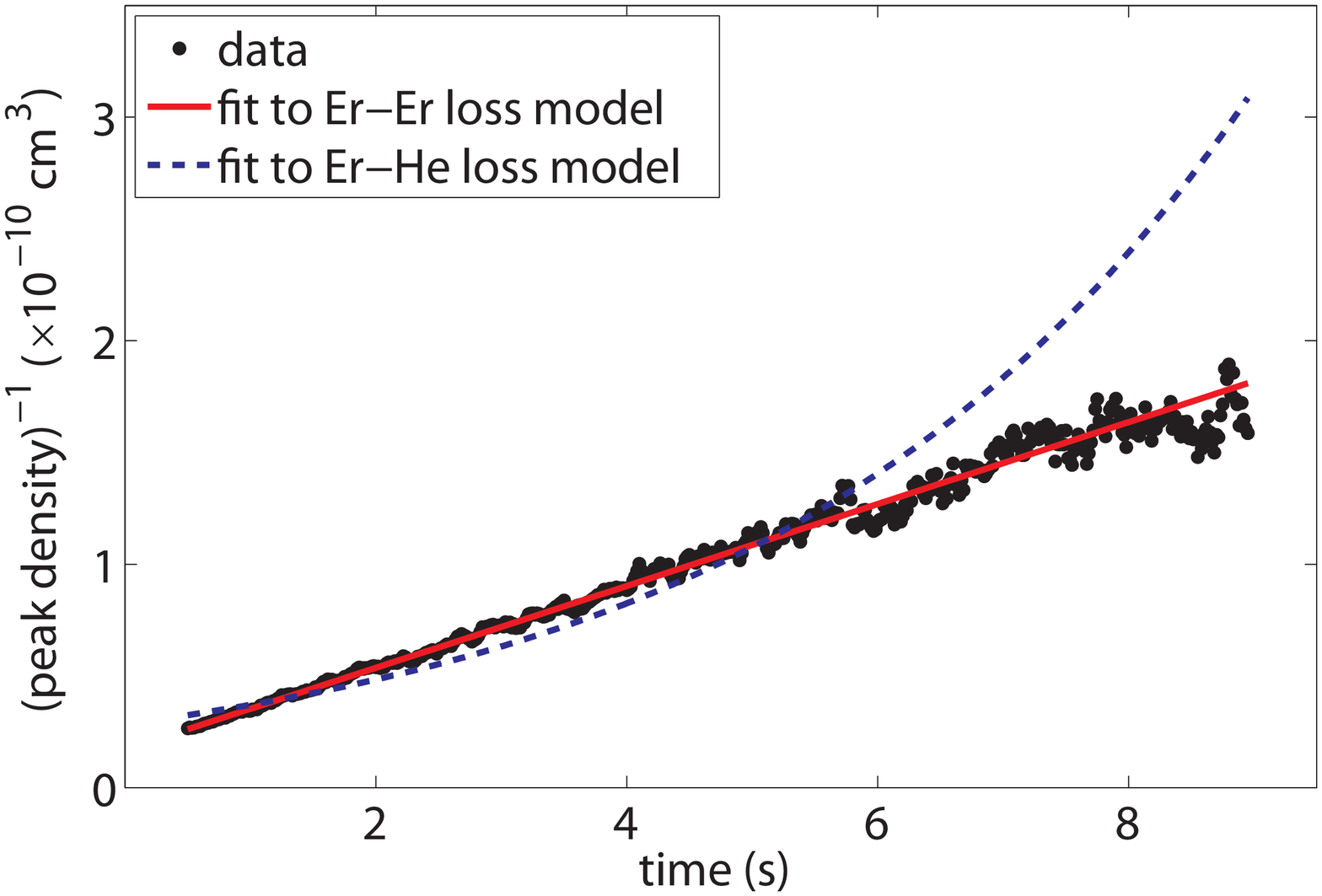}
\caption{\label{fig:Er decay}(color online) Er decay at 530~mK in a $0.99$~T ($4.6$~K) deep magnetic trap after ablation at $t=0$~s.  The vertical axis is the reciprocal of the peak atom density obtained from spectra.  The solid red line is a fit to Eqn.~\ref{2-body decay}. The dashed blue curve is a fit to the exponential decay expected for collisions with a constant He background.  The excellent fit to Eqn.~\ref{2-body decay} ($r=0.998$) indicates that the atom loss is from Er--Er collisions.}
\end{figure}

As noted above, we ensure that the He density is sufficiently low such that neither elastic nor inelastic RE--He collisional loss is observed (see Fig.~\ref{fig:Er decay}).  The trap loss is then determined by the rate equation:
\begin{equation}\label{rate eqn}
\dot{n}(\vec{r}, t)=-[f_{evap}(E_{trap},T) g_{el}+g_{in}] n(\vec{r}, t)^2,
\end{equation}
where $n$ is the local density of trapped atoms, and $g_{el}$ and $g_{in}$ are the rate constants for elastic and inelastic atom--atom collisions.  The function $f_{evap}$ is the fraction of elastic collisions at temperature $T$ that are energetic enough to produce atoms with energy above the trap depth $E_{trap}$ such that the atoms will adsorb on the cold cell walls and be lost from the trap.  In our experiments $T$ is low enough such that $f_{evap} < 1\%$ \cite{Ketterle96}, and thus elastic collisions do not significantly contribute to atom loss.  Ignoring the first term in Eqn.~\ref{rate eqn}, we solve for $n(\vec{r}, t)$, spatially integrate over the trap distribution, and take the reciprocal to reach the simple two-body decay result:
\begin{equation}\label{2-body decay}
\frac{1}{n_0(t)} \equiv \frac{1}{n(r=0,t)} = \frac{1}{n_0(t=0)} + \frac{g_{in}t}{8}.
\end{equation}
Plotting $n_0^{-1}$ versus time yields a straight line of slope $g_{in}/8$.  Data for Er decay is plotted in this manner in Fig.~\ref{fig:Er decay} and fit to Eqn.~\ref{2-body decay}.  Additionally, a combined fit with free parameters for Er--Er and Er--He collisional loss processes yields a Er--He decay rate consistent with zero, and therefore we conclude that the loss is indeed due to Er--Er collisions.


Fits of atom loss to Eqn.~\ref{2-body decay} yield $g_{in}$ to be $1.5\pm0.2\times 10^{-10}$~{\cmps} for Er and $5.7\pm1.5\times 10^{-11}$~{\cmps} for Tm, with accuracy limited by the density calibration determined from spectra.  Both rates are significantly higher than inelastic rates observed for highly magnetic $S$-state atoms such as Cr, Eu, Mn, and Mo \cite{Weinstein02etal, Nguyen07, Kim97etal, Harris07etal, HancoxMo05etal}.  The spin relaxation rate constants for these species were measured in similar magnetic traps at similar temperatures and found to be $\lesssim10^{-12}$~{\cmps}, consistent with the magnetic dipole-dipole interaction \cite{Pavlovic05etal, SuleimanovUnpub} described by
\begin{equation}\label{eq:MDDI}
V_{dipole}(\vec{r}) = \frac{\mu_0}{4 \pi} \frac{\mu^2}{r^3} [( \vec{\mathbf{J}}_1 \cdotp \vec{\mathbf{J}}_2) - 3 (\vec{\mathbf{J}}_1 \cdotp \hat{r}) (\vec{\mathbf{J}}_2 \cdotp \hat{r})].
\end{equation}
where $\mu$ is the magnetic moment and $r$ is the distance between two atoms with angular momenta $\vec{\mathbf{J}}_1$ and $\vec{\mathbf{J}}_2$, respectively.  The spin relaxation rate constant is dependent on the specific form of the interatomic potential, however the general $\mu^4$ scaling implied by Eqn.~\ref{eq:MDDI} provides a relation between dipole-induced inelastic loss rates for atoms of similar electronic structure.  Eu ($\mu=7~\mu_B$) and Mn ($\mu=5~\mu_B$) in particlar have submerged-shell character similar to Er ($\mu=7~\mu_B$) and Tm ($\mu=4~\mu_B$) \cite{CrMoOpenShellLoss}.  Scaling the cross sections measured for Eu \cite{Kim97etal} and Mn \cite{HancoxMo05etal} by $\mu^4$ and averaging yields $g_{in}=3.4\times10^{-13}$~{\cmps} for Er and  $g_{in}=3.5\times10^{-14}$~{\cmps} for Tm.  The observed inelastic rate constants for Er and Tm in our experiments are 2--3 orders of magnitude larger than these scaled dipolar values, inconsistent with the dipolar loss model and implying another loss mechanism.

A significant fraction of atoms ($>20\%$) in our experiments have $m_J\neq J$, as determined by observing $\Delta m_J=0$ transitions on the $415.2$~nm ($J=6\rightarrow5$) line of Er.  Two-body electronic spin exchange collisions will tend to purify the atomic ensemble towards the $m_J=J$ state, but the stability of spectral features with time implies that this is not the case.  In addition, since such collisions conserve the total $m_J$, they cannot cause loss to untrapped states without also populating more strongly-trapped states, which would cause an unobserved net increase in absorption.  Nuclear spin exchange could lead to trap loss, but observed rates for this process in other submerged-shell atoms with only $I>0$ isotopes have shown it to be much slower than the loss observed here \cite{Harris07etal,Kim97etal}.  Hence the observed loss is spin relaxation to untrapped states.  In our analysis, we assume $g_{in}$ to be the same for all pairs of atoms of any $m_J$.

For spin relaxation collisions resulting in a final state with $m_J>0$, relaxation may not lead immediately to trap loss.  In that case, the $g_{in}$ deduced from loss may be smaller than the true spin relaxation collision rate constant, which we will call $g_{sr}$.  Calculations for collisions between He and $L\neq0$ atoms such as Tm and O yield larger rates for $\Delta m_J=\pm 1,2$ transitions than for other transitions, creating effective selection rules \cite{Krems03, BuchachenkoRate06etal}.  Although RE--RE sytems are not theoretically well-understood, if such selection rules held in the case of Er, the $m_J=J=6$ state would on average require several inelastic collisions to reach an untrapped state, contributing to the nonzero $m_J\neq J$ state population noted above.  In addition, collisional energy can promote inelastically colliding atoms to higher $m_J$ states and inhibit loss.  These thermal excitations are suppressed for $g_J \mu_B B \gg kT$, however this condition fails near the trap center where $B=0$.  Considering both these effects, the observed stability of the spectrum suggests that the $m_J$ state distribution achieves a slowly-varying balance between loss and excitation.  We confirmed this model with simulations of inelastic decay, including thermal excitations and exploring a range of initial $m_J$ state distributions and selection rules.  The simulations suggest a ratio $g_{sr}/g_{in}$ of $2.0~^{+1.0}_{-0.5}$.



Currently there exist no theoretical predictions for RE--RE spin relaxation rates due to the complexity of RE electronic structure; however, one reasonable hypothesis to explain rapid spin relaxation of Er and Tm is that it is induced by electrostatic anisotropy, as is observed in anisotropic outer-shell systems.  Experiments with metastable $^3P_2$ states of Ca and Yb have measured inelastic collision rate constants greater than $10^{-11}$~{\cmps} \cite{Hansen06, Yamaguchi08etal}, nearly as large as the Ca*--Ca* and Yb*--Yb* elastic rate constants.  These inelastic rates are similar to those we observe here for Er and Tm atom--atom collisions, suggestive of a complete lack of suppression of electrostatic anisotropy-driven spin relaxation and in contrast to the dramatic suppression of ${>10^4}$ observed for collisions with He.  In the RE--He case, the Born-Oppenheimer potentials corresponding to different projections of the electronic angular momentum onto the internuclear axis are nearly degenerate \cite{BuchachenkoPot06}, which leads to a low probability for reorientation of the magnetic moment.  The RE--RE interaction potentials are much deeper than those of the RE--He system \cite{Zhang07, Buchachenko07}, and so it is possible that electrostatic anisotropy has a similarly stronger effect.


In conclusion, we have measured the loss rate constants for inelastic Er--Er and Tm--Tm collisions and found them to be large.  For comparison, the maximum elastic cross section $\sigma_{el}$ in the absence of shape resonances can be derived from the well-known unitarity limit \cite{ShankarBook}.  Using the $C_6$ constant calculated for the Yb--Yb system \cite{Zhang07} and assuming elastic collisions between submerged-shell lanthanide RE atoms to be similar, we find the maximum $g_{el} = \sigma_{el} \bar{v} \approx 8\times10^{-10}$~{\cmps} at 500~mK.  Hence the ratio $g_{el}/g_{sr} \lesssim 10$ for both Er and Tm, implying that evaporative cooling of these atoms in a magnetic trap will be highly inefficient \cite{Ketterle96}.  At this temperature we expect ${\approx40}$ partial waves to contribute to collisions, and we note that $g_{sr}$ may be different in the ultracold $s$-wave limit.  However, this limit is rather low for these heavy colliding atoms (10--100~$\mu$K), so the multi-partial wave physics will be applicable over a range of experimental conditions.



The large spin relaxation rates for Er and Tm reported here, along with those reported for Ti \cite{Lu09} and recently measured separately for Dy \cite{NewmanUnpub}, represent significant evidence that the submerged-shell character exhibited by roughly a third of the periodic table and
which is responsible for dramatic suppression effects in atom--He collisions does not imply suppression of electrostatic anisotropy-driven spin relaxation in atom--atom collisions.  As a result, the highly successful method of evaporative cooling in a magnetic trap may remain confined to (isotropic) $S$-state atoms.  In addition, lifetimes for optically trapped atoms in $L\neq0$ states may be short due to spin relaxation unless trapped in the absolute ground state.


\begin{acknowledgments}

We would like to acknowledge helpful discussions with Timur Tscherbul. This work was supported by the NSF under grant number 0757157 and through the Harvard/MIT Center for Ultracold Atoms.

\end{acknowledgments}


\bibliographystyle{apsrev}
\bibliography{reference_database}

\begin{thebibliography}{34}
\bibitem[{CrM()}]{CrMoOpenShellLoss}
\bibinfo{note}{The spin relaxation rate constants for Eu and Mn are consistent
  with the $\mu^4$ cross section scaling within experimental error. The same is
  true between outer-shell Cr and Mo, however these rates are several times
  higher.}
\end{thebibliography}

\end{document}